\documentclass[apj]{emulateapj}

\shorttitle{Globular Cluster Systems}
\shortauthors{Harris}

\begin{document}

\title{A Catalog of Globular Cluster Systems:
What Determines the Size of a Galaxy's Globular Cluster Population?}

\author{William E. Harris}
\affil{Department of Physics \& Astronomy, McMaster University,
  Hamilton ON L8S 4M1, Canada}
\email{harris@physics.mcmaster.ca}

\author{Gretchen L. H. Harris}
\affil{Department of Physics \& Astronomy, University of Waterloo,
  Waterloo ON N2L 3G1, Canada}
\email{glharris@astro.uwaterloo.ca}

\author{Matthew Alessi}
\affil{Department of Physics \& Astronomy, McMaster University,
  Hamilton ON L8S 4M1, Canada}
\email{alessimj@mcmaster.ca}

\clearpage

\begin{abstract}
We present a catalog of 422 galaxies with published
measurements of their globular cluster (GC) populations.  Of these, 248 are
E galaxies, 93 are S0 galaxies, and 81 are spirals or irregulars.
Among various correlations of the total number of GCs with
other global galaxy properties, we find that $N_{GC}$ correlates
well though nonlinearly with the dynamical mass of the galaxy bulge
$M_{dyn} = 4 \sigma_e^2 R_e /G$, where $\sigma_e$ is the central
velocity dispersion and $R_e$ the effective radius of the galaxy light profile.  
We also present updated
versions of the GC specific frequency $S_N$ and specific mass $S_M$ versus host galaxy
luminosity and baryonic mass.  These graphs exhibit the previously known U-shape:
highest $S_N$ or $S_M$ values occur for either dwarfs or supergiants,      
but in the midrange of galaxy size ($10^9 - 10^{10} L_{\odot}$) the GC numbers fall 
along a well defined baseline value of $S_N \simeq 1$ or $S_M = 0.1$,
similar among all galaxy types.
Along with other recent discussions, we suggest that this trend may represent
the effects of feedback, which
systematically inhibited early star formation at either very low
or very high galaxy mass, but which had its minimum effect for intermediate
masses.  Our results strongly reinforce recent proposals that
GC formation efficiency appears to be most nearly proportional
to the galaxy halo mass $M_{halo}$.
The mean ``absolute'' efficiency ratio for GC formation that
we derive from the catalog data is $M_{GCS}/M_{halo} = 6 \times 10^{-5}$.
We suggest that the galaxy-to-galaxy scatter around this mean value
may arise in part because of differences
in the relative timing of GC formation versus field-star formation.
Finally, we find that an excellent empirical predictor of total GC population for galaxies of
all luminosities is $N_{GC} \sim (R_e \sigma_e)^{1.3}$, a result consistent with
Fundamental Plane scaling relations.
\end{abstract}

\keywords{galaxies:  general --
galaxies:  star clusters -- globular clusters: general}

\section{Introduction}

A globular cluster system (GCS) is the ensemble of all such star clusters within
a given galaxy.  The history of GCS studies in the astronomical literature can
properly be said to begin with Shapley's (1918) work on the Milky Way GCS, which
he used to make the first reliable estimate of the distance to the Galactic
center.  Next pioneering steps were taken with reconnaisance of the M31 GCS 
\citep[][among others]{hub32,kro60,vet62} and the other
Local Group galaxies \citep[see][for reviews of this early history]{har79,har91}.
However, it was not until discovery and measurement of the rich GC populations
around the Virgo elliptical galaxies had begun 
\citep{bau55,san68,rac68,han77,har76,str81}
that GCS studies began to emerge as a distinct field.

It is now realized that virtually all galaxies more luminous than 
$\sim 3 \times 10^6 L_{\odot}$ (that is, all but the tiniest dwarfs)
contain old globular clusters, and that these massive,
compact star clusters represent a common thread in the earliest star
formation history in every type of galaxy.  The first
`catalog' of GCSs \citep{har79} listed just 27 galaxies,
all from either the Local Group or the Virgo cluster.  By 1991 the number
had grown to 60 \citep{har91} and a decade later to 73 \citep{har00},
with the sample starting to include galaxies in a wider range of environments.
Other compilations for different purposes were put together by
\citet{bro06}, \citet{pen08}, \citet{spi08}, and \citet{geo10}.

In the past decade, many new surveys
of GCSs for galaxies throughout the nearby universe have taken place.
The relevance of GCS properties to understanding galactic structure
and early evolution is becoming increasingly apparent, so the
construction of a complete new catalog is well justified.
A new list may reveal large-scale trends of GCS properties with galaxy
type or environment, and may also provide a springboard for
designing new studies.
Perhaps the most basic question, and one that dates back decades, is simply to ask
what determines the total population of GCs in a galaxy.  The total GC
population size, $N_{GC}$, must relate to
the GC formation efficiency relative to the field-star population
as well as to the later dynamical evolution of the system.
In this paper, we address these questions by using a newly constructed
GCS catalog to search for correlations of cluster population size
with several other global properties of their host galaxies.

\section{The Data Sample and a GCS Catalog}

We have carried out an extensive literature search
to find published studies of galaxies that, at a minimum,
give some useful, quantitative information
for the total number $N_{GC}$ of its globular clusters.  Much of this material
now comes from a few recent
major surveys that have the distinct advantage of being internally
homogeneous, such as for the Virgo cluster galaxies \citep{pen06},
the Fornax cluster \citep{vil10}, nearby dwarf
galaxies \citep{lot04,geo08,geo09}, nearby E and S0 galaxies
\citep{kun01a,kun01b,lar01}, and supergiant E galaxies 
\citep{bla97,bla99,har06,har09}.
The Hubble Space Telescope (HST) cameras including WFPC2, ACS, and most recently WFC3 have
provided a powerful stimulus for the imaging and photometry that
such surveys depend on.

\begin{figure}
\plotone{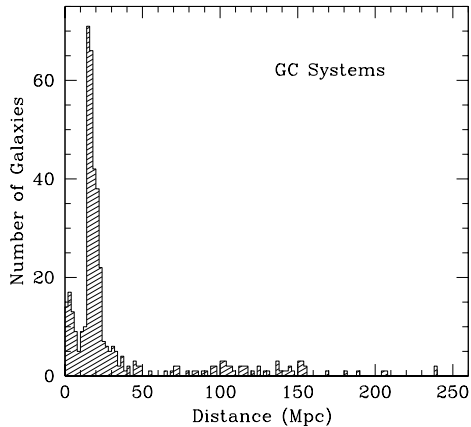}
\caption{Distribution by distance of the 417 galaxies in the catalog.
For comparison, the Virgo cluster is at $D=16$ Mpc and the Coma
cluster at $D= 100$ Mpc.
}
\label{hist_distance}
\end{figure}

However, dozens of other GCS studies exist that are
widely scattered through the literature and we have searched thoroughly for
these as well.
In total we have extracted GCS data from 112 papers published 
up to December 2012.  These yield $N_{GC}$ measurements for 
422 individual galaxies, a dramatic increase over previous compilations.  

The distribution by distance for the galaxies in the catalog
is shown in Figure 1.  Here, the predominance of targets in the 
Virgo-to-Fornax distance range ($15 - 25$ Mpc) is obvious, but the numbers
of more remote galaxies are gradually increasing as new work goes on.  In general, the HST cameras
can reach GCSs reliably for $D \lesssim 200$ Mpc (about twice the Coma
cluster distance), the necessary exposure times becoming quite large beyond that
\citep[cf.][]{har06}.  By comparison, GCS measurement through
ground-based imaging becomes difficult for
$D \gtrsim 40$ Mpc even with 8-meter-class telescopes, 
an empirical limit that partly determines the steep dropoff of targets beyond that
distance.

The complete catalog with literature sources is included in the published
electronic version of this paper, and can also be obtained
from the website of the first author at
http://physwww.mcmaster.ca/$\sim $harris/Databases.html.

In addition to $N_{GC}$, we include some selected 
observational parameters describing the luminosities, masses, and scale sizes 
of the galaxies  and that are
available for most of the objects in the list.  The full catalog contains the
following information:
\begin{itemize}
\item{} Galaxy identification (from more than one catalog source if appropriate).
\item{} Right ascension and declination (J2000).
\item{} Morphological type.
\item{} Foreground absorption $A_V$, from NED (NASA/IPAC Extragalactic Database).  
\item{} Distance $D$, primarily from the raw data in NED.
For relatively nearby galaxies,
wherever possible we adopt $D-$values measured from primary standard candles
based on resolved stellar populations (Cepheids, red-giant-branch-tip stars,
planetary nebulae luminosity functions, Mira stars, RR Lyrae stars).  
Surface brightness fluctuations (SBF) measured from integrated light are 
also used as a standard candle.  For each such galaxy  
we adopt the average of the 
most recent individual measurements of those six primary methods 
(We emphasize that our adopted values are {\sl not} the averages given
in NED).  For some slightly more distant galaxies for which these
primary indicators are not available,
we use recent determinations from the Tully-Fisher
relation as listed in NED.  For still more distant systems ($D \gtrsim 30$ Mpc) we 
use Hubble's law with $H_0 = 70$ km s$^{-1}$ Mpc$^{-1}$ and with the
galaxy radial velocity corrected to the CMB reference frame.
\item{} Absolute visual magnitude $M_V^T$, calculated from the distance
modulus and the integrated magnitude $V_T^0$ if available
from NED.  In other cases where a total $V$ magnitude was unavailable
we have used a blue magnitude $B^T$ and integrated $(B-V)$
color taken from the HyperLeda database.  
\item{} Absolute near-infrared magnitude $M_K^T$, calculated from the distance modulus
and the integrated $K$ magnitude from 2MASS.  We use here the 2MASS $K(ext)$ magnitude
for each galaxy, a quantity which is available for 82\% of the galaxies in our
catalog.  Although an alternate and perhaps preferable choice would be the
frequently used $K_s$ band, this is unavailable for most of our galaxies.
\item{} Total number of globular clusters in the galaxy, $N_{GC}$.
\item{} Stellar velocity dispersion $\sigma_e$.  This spectroscopic quantity is dominated
by the bright inner part of the galaxy and in most cases represents the
velocity dispersion of the bulge light.  Where possible we have taken the homogeneous
$\sigma_e$ values given by \citet{gul09}, \citet{mce95}, and \citet{mcc12}.
Otherwise, we use $\sigma_e$ as compiled in HyperLeda.
In total, $\sigma_e$ measurements are available for 65\% of the galaxies
in the catalog.
\item{} Effective radius $R_e$ enclosing half the total galaxy light, taken
from NED or (secondarily) HyperLeda.  Here, in the interests of the best possible
combination of homogeneity and completeness  we use only radii measured through
optical photometry:  primarily $V$ whenever available, and secondarily other nearby bands
including $g, r$, or $B$.  We do not use any values measured through infrared or near-ultraviolet
bands, since these give systematically different $R_e$ from optical bandpasses.
In total, optically based $R_e$ values are available for 81\% of the galaxies.
\item{} Dynamical mass $M_{dyn}$ of the galaxy, calculated from $\sigma_e$
and $R_e$ as described in Section 3 below.
\item{} Total stellar mass $M_{GCS}$ contained in the entire GC population of
the galaxy, calculated as described below in Section 3.3.
\item{} Measured mass of the central supermassive black hole (SMBH), a quantity
of special interest although it
is currently available for only 11\% of the galaxies in our GCS
list.  SMBH data are taken from \citet{gul09}, \citet{mcc12}, \citet{gra08},
and other sources listed in \citet{har11} and \citet{har13}.
\end{itemize}

\noindent \emph{Comments:}

(1) The only quantities available for \emph{all}
the galaxies in the catalog are $N_{GC}$, galaxy type, and luminosity $M_V^T$.  In principle, a total magnitude
obtained from a near-infrared band such as $z, I$, or $K$ is a better  
photometric proxy for total stellar {\sl mass} than are optical bands.  Internally 
homogeneous near-IR luminosities are available
for subsets of the data \citep[for example, the Virgo and Fornax cluster surveys; see][hereafter P08, V10]{pen08, vil10},
but at present we place most of our reliance on absolute visual magnitudes because
these are available for all of the targets, and allow us to compare various results
readily with earlier work.  In the discussion below, we also present some correlations
with $M_K$ from the 2MASS $K(ext)$ data, but these generally do not exhibit 
any smaller scatter than the ones using $M_V^T$.

(2) The velocity dispersion $\sigma_e$ is a key quantity in many studies of the
``Fundamental Plane'' (FP) of early-type galaxies \citep[e.g.][among many others]{djo87,all09,gra12}.
It is a critical element to dynamical estimates of galaxy mass (see below), and is
a valuable indicator of the depth of the galaxy's potential well \citep{loe03,sha06}.
An additional advantage to including $\sigma_e$  here is that it appears to be stable
with time for early-type galaxies over a significant range of redshift,
as would be the case if large galaxies form their major core light at
high redshift and then evolve later by inside-out growth mainly through minor mergers
\citep{bez09,bez11,tir11,pat12}.
The scale radius $R_e$, by contrast, is expected to grow with time \citep[e.g.][]{pap12,sha13}.

(3) The population size $N_{GC}$ is only a simple first-order gauge of a GCS.
Other GCS characteristics that are of strong interest include the
GCS {\sl radial profile}, the GC {\sl luminosity distribution} (GCLF),
and the {\sl metallicity distribution} (MDF).  The MDF in particular -- 
in most cases measured through broadband color indices -- 
has stimulated much ongoing discussion of the cluster formation process 
\citep[e.g.][among many others]{ash92,lar01,pen06,har09,spi08,mie10}.
Many of the sources listed in our bibliography discuss these other characteristics, but a 
comprehensive analysis of them
extends far beyond the goals of our present study.
For each galaxy we have selected the studies that gave the best
estimates of the total GC population, not necessarily the best
analysis of the MDF or other characteristics.

(4) It should be emphasized that the $N_{GC}$ values collected here are of
greatly differing internal uncertainty from one galaxy to the next
and thus certainly do not make up a homogeneous list.
Ideally, $N_{GC}$ should be determined from imaging that is both
deep enough to reach nearly to the faint limit of the GCLF, and also
wide-field enough to cover the full radial extent of the GCS as
well as to determine the background contamination level accurately.
These twin conditions are rarely met.  The most commonly used technique
is to obtain GC photometry with a limiting magnitude near the ``turnover'' or
peak point of the GCLF and then fit a standard Gaussian GCLF shape to
predict the total over all magnitudes.  If (as is usually
the case) the field coverage does not sample the entire halo of the target
galaxy, then some outward extrapolation of the GCS radial profile beyond
the heavily populated inner regions is
also needed to estimate the total over all galactocentric radii.
Good recent examples of the standard techniques can be found, for example,
in \citet{jor07}, P08, \citet{you12}, and references cited there.
In all cases the GC population
totals are simply the best attempts, with the available imaging data, to
extrapolate over the full luminosity range and spatial range needed.
Inspection of the data catalog will show that the quoted relative uncertainties
$\Delta(N_{GC}) / N_{GC}$ span a wide range:  at best the population total
is known to $\pm10$\% while at worst it may be uncertain by as much as a factor of two.
We return to this point in the later discussion.

(5) Earlier GCS lists were dominated by large E galaxies with rich GC
populations.  The current catalog now reduces many sampling biasses:  it
includes the complete range of galaxy environment, type, and luminosity, from the
smallest dwarfs to the largest supergiants.  Our catalog contains 248
ellipticals, 93 S0's, and 81 spirals or irregulars.  The smallest
one in the list is the dSph KKS-55 at $M_V^T = -11.2$ holding a single GC,
while the largest are cD/BCG supergiants with $M_V^T \simeq -24$ and
holding up to 30,000 GCs each.

(6) We note that for most galaxies in this catalog, we have used the
single literature source that gives the best recent estimate of the total
GC population.  This may not correspond to the best sources for other
purposes, such as discussions of the MDF.  In some cases we have averaged
the results from two or more sources that appear to give comparably good
estimates of $N_{GC}$.  Lastly, in the catalog we have chosen {\sl not} to
list any galaxies for which the estimated $N_{GC}$ was zero or negative.
Such cases include a few very small dwarfs \citep{pen08,geo10} for which
$N_{GC} \leq 0$ after subtraction of field contamination; or a few galaxies
for which the imaging data simply did not reach deep enough to provide a
sensible estimate of GC numbers \citep{kun01a, kun01b}.  An exception is
the Local Group elliptical M32, which we include for historical reasons
though it has no clearly identified clusters.

\begin{figure}
\plotone{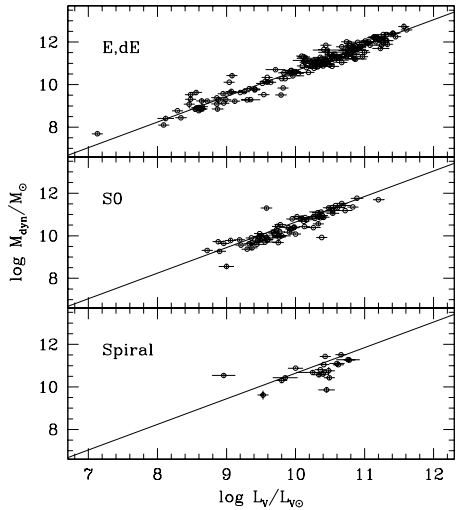}
\caption{`Dynamical mass' $M_{dyn} = 4 R_e \sigma_e^2 / G$ plotted versus 
total visual luminosity $L_V$ for the galaxies in the GCS catalog.
Results are shown separately for the E/dE, S0, and spiral types.
In each panel the diagonal line shows the best-fit linear solution to
the ellipticals, as described in the text.
}
\label{lmass_3panel}
\end{figure}

\begin{figure}
\plotone{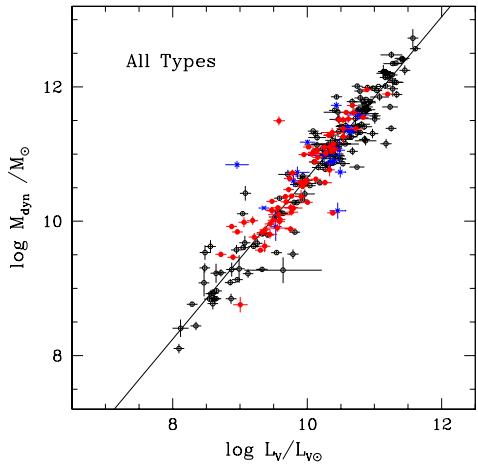}
\caption{$M_{dyn}$ versus 
total visual luminosity $L_V$ for all galaxies in the GCS catalog.
The E/dE systems are plotted as open circles, S0 as red filled symbols,
and S/Irr systems as blue crosses.  An arbitrary offset of +0.2 dex has been applied
to (log $M_{dyn}$) for the S0 points, and +0.3 dex to the S/Irr points
to bring them in line with the E-galaxy solution (see text).
The diagonal line shows the best-fit linear solution to
the ellipticals as in the previous figure.  
}
\label{lmass_all}
\end{figure}

\section{Correlations}

Using the entire database, we next explore some simple correlations of $N_{GC}$ with 
large-scale host galaxy properties such as luminosity, dynamical mass, 
or scale size.  The overall purpose is to use this new and larger dataset to search for \emph{reliable
predictors of GC population size} that can be calculated from the shortlist of simple structural
parameters that are available for most of the galaxies.  In doing so, we concentrate particularly on the
ellipticals in the catalog, for two reasons:  (a) they cover the largest range in
luminosity, from dwarf-spheroidal up to cD supergiants; and (b) they form the largest
and most homogeneous subset of the GC measurements, making up almost 60\% of the
entire catalog.

In general we take pairs of parameters (x,y) in log/log space
and search for linear correlations of the normal form $y = \alpha + \beta x$.
Assume that we have a sample of $n$ measured datapoints $(x_i, y_i)$
with quoted measurement uncertainties $(\sigma_{xi}, \sigma_{yi})$.
Our best-fit slope and zeropoint are determined by minimizing the sum
\begin{equation}
\chi^2 = \sum_{i=1}^n {(y_i - \alpha - \beta (x_i - x_0))^2 \over {(\sigma_{yi}^2 + \epsilon_y^2)
   + \beta^2 (\sigma_{xi}^2 + \epsilon_x^2)} }   
\end{equation}
as in \citet{tre02} and particularly \citet{nov06}.
Here $x_0$ is a suitable mean value over the 
the sample such that $\langle x-x_0 \rangle \simeq 0$
to minimize the covariance between $(\alpha, \beta)$.

The parameters $(\epsilon_x, \epsilon_y)$ are constants that represent any 
additional variances in 
$(x, y)$.  These variances might be due to intrinsic (``cosmic'') scatter built in to the sample
population; or extra measurement uncertainties if the quoted $(\sigma_x, \sigma_y$) are
underestimated; or a combination of both effects.
With only two pairs of measurements $(x_i, y_i)$ in the solution, and without any other
external constraints, in general it will not
be possible to solve independently for both of $\epsilon_x, \epsilon_y$.
Therefore in practice for each solution described below, we set $\epsilon_x = 0$ and vary
$\epsilon_y$ until $\chi_{\nu}^2$, the reduced $\chi^2$ per degree of freedom, 
equals 1.  In the discussion below we refer to $\epsilon_y$ determined from
the solution as the {\sl residual dispersion} for the dependent variable $y$
\citep[see][]{nov06}.

To set up several of the correlation solutions discussed later,
we also calculate 
(a) the total visual luminosity $L_V = 10^{0.4(M_{V \odot} - M_V^T)} L_{V\odot}$
with Solar $M_{V\odot} = 4.83$; and (b) the dynamical mass,
\begin{equation}
M_{dyn} \, = \, {{4 R_e \sigma_e^2} \over {G}}
\end{equation} 
following \citet{wol10}.  Since the luminosity-weighted velocity
dispersion is dominated by light from within $R_e$, and the dark-matter halo
contributes a small fraction of the mass within $R_e$, $M_{dyn}$ is
close to being the baryonic mass of the galactic bulge 
\citep[e.g.][]{tir11,gra12}.  The dynamical mass can be calculated for 61\%
of the galaxies in our catalog, i.e. the ones with measurements of both $\sigma_e$ and $R_e$.

As a preliminary step and a check of our procedures, we show in
Figure \ref{lmass_3panel} the direct
correlation between $M_{dyn}$ and visual luminosity $L_V$.
The data are shown separately for the E, S0, and spiral galaxies (note
that no dwarf irregulars are present here, since they lack any ``bulges''
from which a velocity dispersion can be measured).  It is worth emphasizing
that $L_V$ and $M_{dyn}$ are observationally nearly independent measurements except
insofar as $R_e$ relies on knowing the large-scale light profile of the
galaxy.  For the E galaxies over their entire range, the best-fit linear solution
is listed in Table 2, along with the other correlations to be
presented below.  In the Table, the successive columns give (1) the pair of
parameters (x,y) being fit, (2) the subsample of galaxy types used in the solution,
(3) the number of galaxies in the solution,
(4-6) the sample mean $x_0$, zeropoint $\alpha$, and slope $\beta$, 
(7) the best-fit residual dispersion $\epsilon_y$, and (8) total rms
scatter $\sigma_y$ of the datapoints around the fitted solution.
Throughout Table 2, the luminosities $L_V$ and masses $M_{dyn}$ are
in Solar units.

\begin{figure}
\plotone{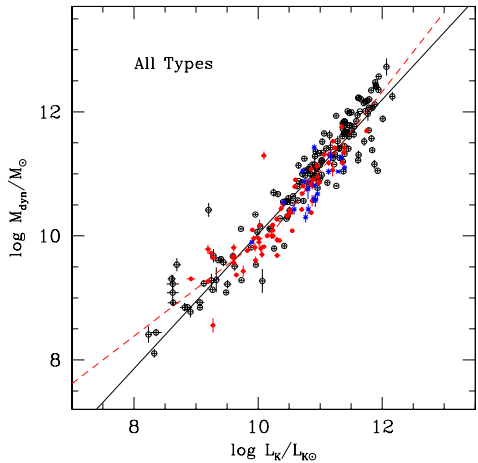}
\caption{$M_{dyn}$ plotted versus 
total $K-$band luminosity $L_K$ for the galaxies in the GCS catalog.
The E/dE systems are plotted as open circles, S0 as red filled symbols,
and S/Irr systems as blue crosses.  
The solid diagonal line shows the best-fit linear solution to
the entire sample, as described in the text, whereas the dashed line
shows the two-part solution over the bright and faint ranges as in Table 2.
}
\label{irmass}
\end{figure}

The result $M_{dyn} \sim L_V^{1.2}$
corresponds to the well known systematic increase in mass-to-light ratio
with galaxy size derived from the Fundamental Plane (FP) of early-type galaxies
\citep[e.g.][]{fab87,jor96,cap06,all09,mag12,gra12}.
Quantitatively, our E galaxy solution  corresponds to 
\begin{equation}
(M/L)_V = 4.406 (L_V / {10^{10} L_{V\odot}})^{0.2}
\end{equation}

The same slope also matches the S0 and
spiral-galaxy trends (lower two panels of Fig.~\ref{lmass_3panel}), although
these disk galaxies fall below the E line by 0.2-0.3 dex in mass (or alternately,
they lie at higher luminosity for a given mass by the same average factor).  We use
the same mass calculation formula (Eq.~(2)) for all types, though it is not entirely
clear that pure-spheroid (elliptical) systems and disk systems should behave identically
or should have the same $M/L$.  
The measured scatter of $\pm 0.27$ dex rms in log $M_{dyn}$ about the best-fit
line is also encouragingly small, given that the luminosity and calculated
mass are derived from a wide variety of observational sources for
$V^T, \sigma_e$, and $R_e$ and are unavoidably a somewhat heterogeneous
sample.  As will be seen below, we find very similar scatters for most
of our other correlations.
In Figure \ref{lmass_all}, we show all galaxy types together where an offset
of 0.2 dex to the calculated mass has been applied to the S0 systems to
bring them in line with the ellipticals, and an offset of 0.3 dex to the S/Irr systems.

\begin{figure}
\plotone{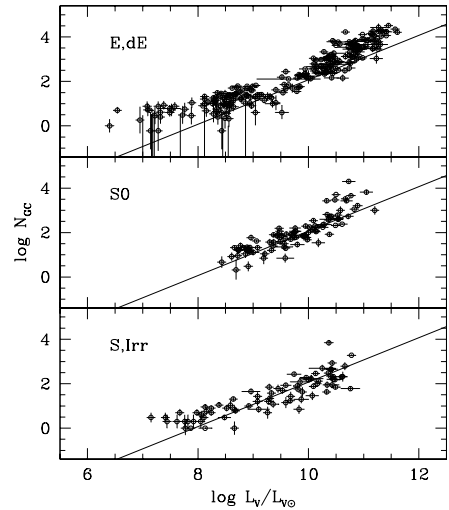}
\caption{\emph{Top panel:} Total number of globular clusters $N_{GC}$ plotted versus the
visual luminosity of the host galaxy, for elliptical galaxies. 
\emph{Middle panel:} The same plot for S0 galaxies. 
\emph{Lower panel:}  The same plot for spiral and irregular galaxies.
In all three panels, the sloped line denotes a specific frequency $S_N \equiv 1$.
}
\label{lum_3panel}
\end{figure}

In Figure \ref{irmass} we show a similar correlation for the $K-$band luminosity
versus $M_{dyn}$.  Plotting the three types of galaxies (E, S0, S/Irr) separately
indicates little or no zeropoint difference, so we show all three combined without offsets.
Here we adopt (log $L_K/L_{K\odot}$) $= 0.4 (3.33 - M_K)$.
The best-fit solution over all luminosities (Table 2) gives $M_{dyn} \sim L_K^{1.085 \pm 0.020}$
and has a very similar scatter of $\pm 0.26$ dex.
However, unlike $L_V$ (Fig.~\ref{lmass_all}), the relation exhibits a noticeable nonlinearity
where the brighter galaxies follow a steeper slope.  Splitting the data at $M_K = -22.4$
(or log $L_K/L_{K\odot}$ = 10.3) gives the two additional solutions listed in Table 2,
where $M_{dyn} \sim L_K^{0.765 \pm 0.060}$ for fainter galaxies and
$M_{dyn} \sim L_K^{1.260 \pm 0.048}$ for brighter ones.

In summary, both $V$ and $K$ luminosities in our catalog can act as similarly
precise indicators of galaxy dynamical mass.   Where necessary, in the discussion
below we choose to use $L_V$ because it is available for the entire catalog of galaxies.

\subsection{Cluster Population and Galaxy Scale Parameters}

\begin{figure}
\plotone{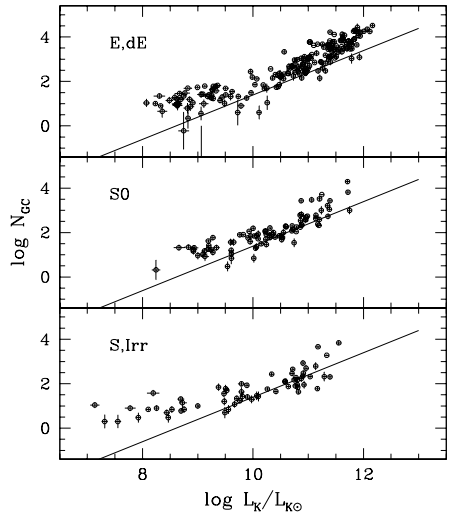}
\caption{\emph{Top panel:} Total number of globular clusters $N_{GC}$ plotted versus the
$K-$band luminosity of the host galaxy, for elliptical galaxies. 
\emph{Middle panel:} The same plot for S0 galaxies. 
\emph{Lower panel:}  The same plot for spiral and irregular galaxies.
In all three panels, the sloped line denotes a specific frequency $S_N \equiv 1$.
}
\label{lumir_3panel}
\end{figure}

\begin{figure}
\plotone{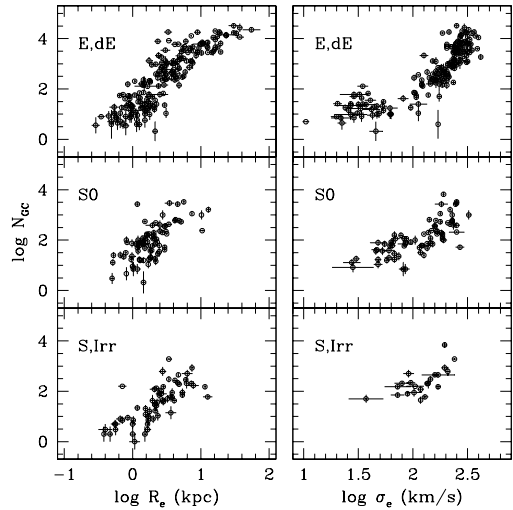}
\caption{\emph{Left panels:} Total number of globular clusters $N_{GC}$ plotted versus 
the effective radius $R_e$ of the host galaxy.  Elliptical, S0, and S/Irr galaxy types
are plotted separately.
\emph{Right panels:} Total number of globular clusters $N_{GC}$ plotted versus 
the bulge velocity dispersion of the host galaxy. 
}
\label{resig_3panel}
\end{figure}

\begin{figure}
\plotone{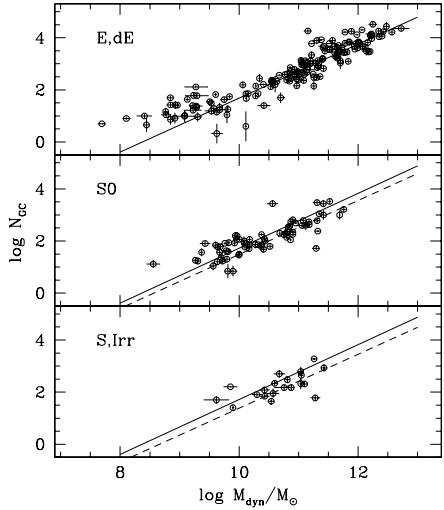}
\caption{Correlation of GC population size $N_{GC}$ versus
the dynamical mass $M_{dyn} = 4 R_e \sigma_e^2 / G$.
The data are shown separately for ellipticals, S0's, and spirals as in
previous figures.  In each panel the {\emph solid diagonal line} shows
the best-fit solution for the luminous E galaxies as discussed in the
text, i.e. excluding the dwarfs.  In the middle panel, the dashed line is
the E-galaxy solution shifted downward by 0.2 dex, while in the lower panel
the dashed line is the E line shifted downward by 0.3 dex (see text).
}
\label{mass_3panel}
\end{figure}

\begin{figure}
\plotone{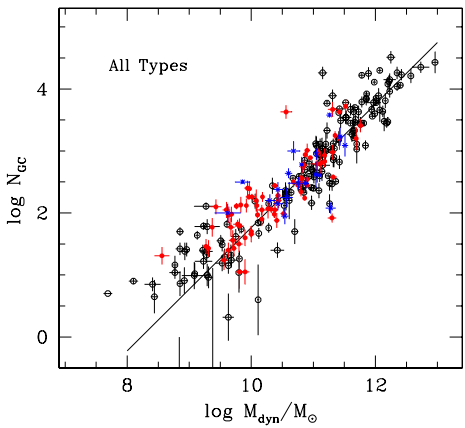}
\caption{Correlation of GC population size $N_{GC}$ versus
the dynamical mass $M_{dyn} = 4 R_e \cdot \sigma_e^2 / G$.
E galaxies are plotted as open circles, S0's as solid red circles, and
spirals as blue crosses.  The $N-$ values for the S0 and spiral types
have been normalized to the E-galaxy level as described in the text.
The {\emph solid diagonal line} shows
the best-fit solution for all galaxies with $M_{dyn} > 10^{10} M_{\odot}$.
}
\label{mass_all}
\end{figure}

We next plot total cluster population $N_{GC}$ versus the other listed scaling
parameters:  $L_V$, $\sigma_e$, $R_e$, and $M_{dyn}$.  
These are displayed in Figures \ref{lum_3panel} to \ref{mass_all},
and selected best-fit solutions are listed
in Table 2.  Comparison with
the remaining quantity in the list, the SMBH mass, 
is a special topic that has been the subject of
discussion in several recent papers including \citet{spi09} (hereafter S09); 
\citet{bur10,har11,sny11,sad12,rho12,har13},
and will not be repeated here.

The most easily observable relation is between $N_{GC}$ and the galaxy $V$ luminosity, which
was also historically the first to be discussed in the literature
\citep{jas57,han77,har79,har81}.
The data for all the systems in the
current catalog are shown in Figure \ref{lum_3panel}.  As has been found in 
all earlier discussions, $N_{GC}$ increases very roughly in direct proportion
to host galaxy luminosity, but obvious systematic deviations occur at both the
high- and low-luminosity ends of the scale, and between E/S0 systems and
S/Irr ones.

In Figure \ref{lumir_3panel}, we show the correlations between $N_{GC}$
and near-infrared luminosity $M_K$.  The pattern is very much the same,
and the scatter quite similar to the previous figure.  Since the $N_{GC} - M_K$
graph appears to give much the same information, and $M_V^T$ is available for a
larger sample of galaxies, we stick primarily with the use of
the visual-luminosity data in the following discussion.

\begin{figure}
\plotone{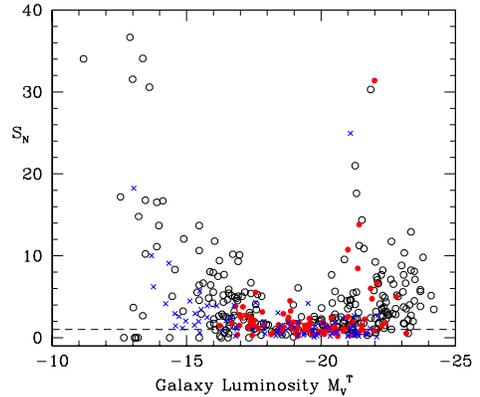}
\caption{Specific frequency $S_N$ versus the absolute visual magnitude 
$M_V^T$ of the host galaxy.  E and dE galaxies are plotted as open circles,
S0 systems as solid red circles, and spirals or irregulars as blue crosses.
The horizontal line at bottom shows $S_N = 1$.
}
\label{sn}
\end{figure}

The connection between galaxy size and GC population is most often presented
in terms of the specific frequency \citep{har81},
\begin{equation}
S_N \, \equiv \, N_{GC} \times 10^{0.4(M_V^T + 15)} \, = \, 
      (8.51 \times 10^7) \, {N_{GC} \over {L_V / L_{V\odot}} } \, .
\end{equation}
In the $K-$band, if we adopt a typical color index $(V-K) \simeq 3.2$
for E galaxies and bulges \citep[e.g.][]{mic05}, then the equivalent
relation is
\begin{equation}
S_N \, \equiv \, N_{GC} \times 10^{0.4(M_V^T + 18.2)} \, = \, 
      (4.1 \times 10^8) \, {N_{GC} \over {L_K / L_{K\odot}} } \, .
\end{equation}
The trend of $S_N$ versus $M_V^T$ is shown in Figure \ref{sn}, for the
three subsets of data combined.
The characteristic U-shaped distribution is the most prominent feature
of the diagram:  the intermediate-luminosity
galaxies form a rather tight grouping in a ``valley'' around $S_N \simeq 1$ and the
dwarfs and supergiants at opposite ends show much larger scatter and higher mean
$S_N$.  This distribution was apparent
even in the first discussion of specific frequency from a sample more than
an order of magnitude smaller \citep[see Fig.~4 of][]{har81}.
Two notable recent versions with extensive discussions are given by
P08 and \citet{geo10} (hereafter G10), which build on the
newer surveys and particularly fill in the lower-luminosity range more extensively than before.

In general the highest specific frequencies are found either in some E supergiants
(particularly the cD or BCG giants) or in dwarf spheroidals and nucleated
dE,N galaxies.\footnote{At the top end, the anomalously high$-S_N$ values seen
in Fig.~\ref{sn} for
four S0 or S galaxies are those for NGC 6041A, UGC 3274, A2152-2, and IC 3651.
All are distant and luminous systems in rich clusters of galaxies, and close
inspection of images suggests that they may simply be misclassified E/cD systems.}
S0 and disk galaxies have systematically lower $S_N$ than ellipticals,
field E's have lower $S_N$ than ones in rich clusters of galaxies, and
dE's in denser environments favor higher $S_N$ 
\citep[e.g.][and P08]{han77,har81,vdb82,vdb00,dur96,har01,bro06}.

In most galaxies there are two clearly identifiable subsets of GCs
that separate out by color or metallicity:  the blue (metal-poor)
population and the red (metal-rich) ones \citep[e.g.][]{pen06,har09,mie10}.
The ratio N(red)/N(blue) is dependent on galaxy size, with
GC populations in lower-luminosity galaxies of all types progressively
more dominated by the metal-poor component \citep{pen06}.
The blue GCs are likely to be the remnants of the very earliest 
star-forming stages of hierarchical merging, emerging out of the gas-rich and metal-poor
protogalactic dwarfs \citep[e.g.][and P08]{bur01,moo06}, or added later by
accretion of low-mass metal-poor satellites.  
With photometric data of sufficient quality, it is also possible to 
define $S_N$ or the GC mass fraction for red and blue types separately
\citep[e.g.][]{rho05,rho07,bro06,spi08,for09}.  However,
many of the galaxies in our catalog do not have sufficient photometric data to evaluate the blue/red ratios
accurately, and we do not pursue this question here.  New photometric
data aimed at obtaining high-quality blue/red population ratios for more galaxies
would be of great interest.  In particular, it would be important to know how
much of the scatter around the mean $S_N$ relation at a given galaxy luminosity
might be due solely to differences in the relative number of blue GCs (and thus
the efficiency of cluster formation at the earliest stages).
The references cited above should be seen for more complete discussion.

\subsection{Other Correlations for $N_{GC}$}

Going beyond specific frequency, we have explored more general correlations
of $N_{GC}$ versus combinations of scale size and velocity dispersion.
We might, for example, hope
to find choices which would more nearly linearize the trend
of $N_{GC}$ over the \emph{entire} galaxy luminosity range from dwarfs
to supergiants.

Neither $R_e$ nor $\sigma_e$ by itself is a good predictor of GC population.
As seen in Fig.~\ref{resig_3panel}, $N_{GC}$ versus those quantities exhibits quite
a lot of scatter and behaves nonlinearly.  The exception here is $N(R_e)$ for the
spiral and irregular galaxies, which yields a roughly useful scaling in the
cases where $\sigma_e$ is not available (see discussion below).

$N_{GC}$ versus $M_{dyn}$ is plotted in Fig.~\ref{mass_3panel} for the three
galaxy types, and in Fig.~\ref{mass_all} for all galaxies combined.
Similar trends of GCS numbers versus mass
are also discussed by P08, S09, and G10.  The
significant difference compared with our work is that these previous studies employed 
{\sl photometrically}
determined masses (e.g. the combination of a near-infrared luminosity and an
assumed mass-to-light ratio), whereas we use $M_{dyn}$ which stands
independently of photometric indicators.  

For galaxy masses $> 10^{10} M_{\odot}$,
we find that $N_{GC} \sim M_{dyn}^{1.04 \pm 0.03}$. That is, GCS population
increases  in almost exactly direct proportion to galaxy mass.  For the smaller galaxies, the scaling
is much shallower at $N_{GC} \sim M_{dyn}^{0.4}$ and these also exhibit larger
scatter (see particularly Fig.~\ref{mass_all}).
We find as well that the S0-type galaxies
lie below the ellipticals by $\Delta$ log $N_{GC} \simeq -0.2$ dex, while the spiral types
fall even further below the ellipticals by -0.3 dex (again, no irregulars appear in this
plot).  In short, {\sl if} the same definition of $M_{dyn}$ is valid for disk galaxies
and ellipticals (cf. the caveats mentioned earlier), then
disk galaxies have fewer clusters per unit bulge mass than do
ellipticals, by factors of 1.5 to 2.  This point is discussed more extensively
by G10.  To plot up Fig.~\ref{mass_all} we have applied these offsets to the S and S0
types to bring them back to the E/dE line.

\begin{figure}
\plotone{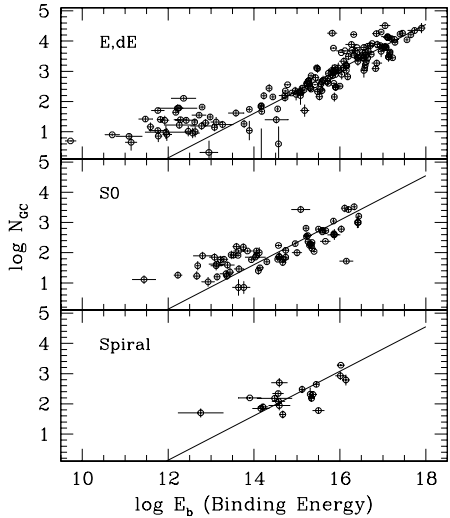}
\caption{Correlation of GC population size $N_{GC}$ versus
the binding energy $E_b \sim  R_e \cdot \sigma_e^4$, as described
in the text.  In each panel the {\emph solid diagonal line} shows
the best-fit solution for the luminous E galaxies,
i.e. excluding the dwarfs.
}
\label{eb}
\end{figure}
               
In Figure \ref{lumir_3panel}, we show $N_{GC}$ now plotted against $K-$band infrared
luminosity.  In principle, if near-IR luminosity is a valid proxy for total
stellar mass, then this graph should reveal the same basic trend as does
Fig.~\ref{mass_all}.
It does show the same trend, but the scatter is similar to the correlations
with $L_V$ and so is not
additionally useful for the present purposes.

In addition to $M_{dyn} \sim R_e \sigma_e^2$, another quantity used occasionally in 
the literature especially for
pressure-supported systems such as star clusters, molecular clouds, or
E galaxies, is the system's
binding energy $E_b \sim M \sigma^2 \sim R_e \sigma_e^4$
\citep[e.g.][]{mcl00,hop07,sny11}.
For completeness we show the correlation of $N_{GC}$ and $E_b$ in
Figure \ref{eb}, where numerically we define 
$E_b = M_{dyn} (\sigma_e/[200 {\rm km~s}^{-1}])^2$.
Once again, the luminous galaxies (log $E_b > 13.5$) form a well defined 
relation close to $N_{GC} \sim E_b^{3/4}$, with total scatter quite similar
to the previous solution between $N_{GC}$ and $M_{dyn}$ (see Table 2).  However, the
dwarf galaxies stand even further off the mean line than before, so
there appears to be no additional advantage to using $E_b$ as a predictor 
of GC population.

\begin{figure}
\plotone{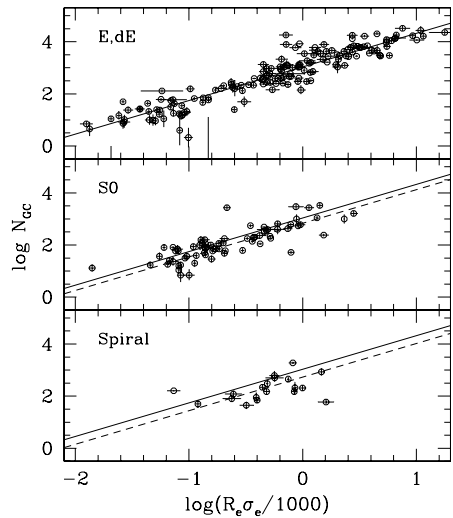}
\caption{GC population size $N_{GC}$ versus
$(R_e \sigma_e)$ as defined in the text.
The data are shown separately for ellipticals, S0's, and spirals as in
previous figures.  In each panel the {\emph solid diagonal line} shows
the best-fit solution for the E galaxies, but now including {\emph both
dwarfs and giants}.  In the second panel, the dashed line shows the E-galaxy
line shifted downward by 0.2 dex, while in the lower
panel the dashed line shows the E-galaxy solution shifted down by 0.3 dex.
}
\label{genfit}
\end{figure}

Going in the opposite direction to a smaller power of $\sigma_e$  has the numerical
effect of reducing its importance and bringing the dwarfs
closer to the giant-galaxy line.  We have explored a range of different
empirical combinations and, as an example, we show the case
for (log $N_{GC}$) against the direct product  
(log $R_e \sigma_e$) in Figure \ref{genfit}.  
This result comes close to giving a nearly linear correlation with
encouragingly low scatter, over the \emph{entire luminosity range}
of galaxies from the smallest dwarfs to the largest supergiants,
a range of 5 orders of magnitude in mass.
In performing the fit we have deleted the five most deviant points (three
dwarfs, two giants), leaving $N=158$ galaxies to determine the solution.
In Fig.~\ref{genfit} the E-galaxy solution is also shown superimposed on
the data for the 72 S0 and 19 spiral systems.  Again,
the E solution adequately matches
the S0's for a -0.2 dex shift in log $N_{GC}$, and matches the spirals for a -0.3 dex shift
(shown as the dashed lines in the lower two panels).  

In brief, we find that \emph{the total globular cluster population of
a galaxy is accurately predicted by the simple product of the galaxy's effective
radius $R_e$ and bulge velocity dispersion $\sigma_e$}.  The specific relation
for the E galaxies is
\begin{equation}
N_{GC} \, = \, (600 \pm 35) \left[ ({R_e \over {10 {\rm kpc}}}) ({\sigma_e \over {100 {\rm km/s}}}) \right]^{1.29 \pm 0.03}
\end{equation}
The same relation can also be used for S0 and spiral types, with the zero-point shifts
given above.

This simple relation is not useful for late-type spiral or irregular galaxies where $\sigma_e$
is not defined or not measurable.  In those cases, a rough but still useful predictor
of $N_{GC}$ appears to be the effective radius $R_e$ alone, as seen in Fig.~\ref{resig_3panel}
(lower left panel).  For these types of galaxies, we find
\begin{equation}
N_{GC} \, = \, (38 \pm 7) (R_e / 2.5 {\rm kpc})
\end{equation}
(listed in its log/log form in Table 2).  The scatter in this case is $\pm 0.53$ dex in
log $N_{GC}$, significantly higher than for the other relations presented above, and
reflecting the intrinsically wide range of GC systems found in star-forming dwarf
galaxies.  
Nevertheless, it should be useful for giving first-order estimates if no other 
recourse is available.

\subsection{Specific Mass and Galaxy Scale Parameters}

The total number of GCs in a galaxy is only a proxy for a more physically
relevant quantity, the total stellar {\sl mass} $M_{GCS}$ contained in all
the GCs.  Ultimately, we would like to know the typical fraction of baryonic mass
or total halo mass taken up by the GCs.   
Here we use the ``specific mass'' defined as a percentage of the previously
calculated dynamical mass of the host galaxy,
\begin{equation}
S_M \, = \, 100 {M_{GCS} \over M_{dyn} } \, .
\end{equation}
This ratio should in principle be similar to the definition $S_M = 100 M_{GCS} / M_{G\star}$
used by P08, since as discussed above, $M_{dyn}$ is
nearly equal to the total bulge stellar mass with minor contributions 
from either dark matter or gas (except for major star-forming systems).
By comparison, G10 use $S_M = 100 M_{GCS} / (M_{\star} + M_{gas})$,
which can be significantly different for either gas-rich dwarfs or
cD-type systems with massive amounts of hot halo gas.
Perhaps more importantly however, our discussion of $S_M$ differs from those of
P08, S09, or G10
in that they used \emph{photometrically estimated} stellar masses $M_{\star}$
versus our dynamical masses.

To define $S_M$ we need to add up the masses of the GCs in a given galaxy,
or equivalently find the mean GC mass.  For studies of GCSs such as the
Virgo Cluster survey (Peng et al.) or ones in very nearby galaxies, it is
possible to obtain a nearly complete census of all the clusters in a given galaxy and explicitly add them up 
one by one.  For most of the galaxies in our catalog, however, we 
must adopt a more broad-brush procedure in hopes of
identifying first-order trends.  What helps considerably is the
empirical fact that the GCLF has a consistent, predictable shape across
all galaxies, which conveniently allows us to scale $N_{GC}$ to $M_{GCS}$ fairly
straightforwardly.  

Usually the GCLF is represented by a simple Gaussian in number of GCs
per unit magnitude interval, $n(M_V)$, with a characteristic peak $\mu_V$
and standard deviation $\sigma_V$.  More detailed analysis suggests that a
slightly asymmetric form such as the ``evolved Schechter function'' developed by
\citet{jor06, jor07} 
is a better match to the data.  However, this is only
a minor concern because the total
mass in the GCS is dominated by the clusters brighter than
the peak (turnover) point of the GCLF; the clusters fainter than the turnover make up
typically only $\simeq20$\% of the total GCS mass. Thus the normal Gaussian-type analytical
approximation where $\sigma_V$ is determined by the bright half of the GCLF
remains quite useful.  

\begin{figure}
\plotone{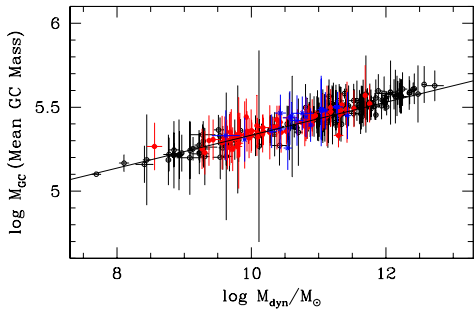}
\caption{Mean globular cluster mass $\langle M_{GC} \rangle$ versus
galaxy mass $M_{dyn}$.  Here, the mean GC mass is defined as $\langle M_{GC} \rangle
= M_{GCS} / N_{GC}$ and the total mass in the globular cluster system is calculated
as described in Section 3.3.  Symbols for the three galaxy types (E, S0, S/Irr) are
as in previous figures.  The diagonal line gives the best-fit relation 
$\langle M_{GC} \rangle \sim M_{dyn}^{0.098}$ (Table 2).
}
\label{meanmass}
\end{figure}

A more important consideration is that $\mu_V$ and
$\sigma_V$ depend on galaxy luminosity, in the sense that the GCLF becomes
broader and brighter for bigger galaxies.  To integrate over the GCLF,
we follow the relations derived by \citet{jor06} and V10,
adopting $\mu_V = -7.4 + 0.04(M_V^T + 21.3)$ and 
$\sigma_V = 1.2 - 0.10 (M_V^T + 21.3)$, where the zeropoints are normalized to
the Milky Way.\footnote{The exception is that we set a lower
bound $\sigma_V(min) = 0.5$.  Inspection of the data from V10 shows that 
the deduced slope $\Delta \sigma / \Delta m = -0.10$ depends heavily on the more luminous
galaxies, whereas for the dwarfs there is mainly a large scatter with no clear trend.}
Finally, to convert GC luminosity to mass, we use a constant $(M/L)_V = 2$
\citep{mcl05}.\footnote{We note that S09 use a constant
$\mu_V$ and $\sigma_V$ to calculate $M_{GCS}$.}  Figure \ref{meanmass} shows the resulting
trend of mean individual GC mass versus galaxy mass $M_{dyn}$ (where now 
$\langle M_{GC} \rangle \equiv  M_{GCS} / N_{GC}$).  The residual scatter around this
relation is simply the visible result of the galaxy-to-galaxy differences in the calculated
$M_{dyn} = f(R_e, \sigma_e)$ for a given galaxy luminosity.  
This graph would be a dispersionless relation if, as in other
papers, we had used galaxy \emph{luminosity} to determine galaxy mass.
The overall trend 
is listed in Table 2 and shown in Figure \ref{meanmass} and is well matched by a single power law,  
$\langle M_{GC} \rangle = (2.26 \times 10^4) M_{dyn}^{0.098}$ in Solar masses.  

The correlation solution for GCS mass versus galaxy mass, for the E galaxies with $L > 10^{10} L_{\odot}$ (Table 2),
yields $M_{GCS} \sim M_{dyn}^{1.16 \pm 0.04}$, 
slightly but significantly steeper than the $N_{GC} \sim M_{dyn}^{1.04}$ dependence found earlier.
The difference is a direct result of the second-order trend for mean GC mass to increase with
host galaxy size.
For the smaller E galaxies, the mean trend is 
$M_{GCS} \sim M_{dyn}^{0.46 \pm 0.10}$, again quite similar to the dependence of $N_{GC}$ on  $M_{dyn}$.

The plot of $S_M$ versus $M_{dyn}$ is shown in log/log form in Figure \ref{sm}.  The overall distribution is
roughly similar to that for $S_N$ (Fig.~\ref{sn}), though with less scatter at either the high-luminosity
or low-luminosity ends.  This reduced scatter is partly a result of our use of dynamical masses rather
than photometric masses for the galaxies (for example, two giant ellipticals may have the same luminosities,
but if one of them is a cD-type or BCG, it will usually have a larger effective radius or central velocity
dispersion and thus a higher dynamical mass).   

The practical penalty for using $S_M$ instead of $S_N$ is that we cannot strictly include as many datapoints
because we need to have both $R_e$ and $\sigma_e$ to determine $M_{dyn}$.
We therefore supplement the dynamical data by adding in photometrically calculated masses from the known conversion
between $L_V$ and $M_{dyn}$, from Eq.~(3) and Table 2.  These secondary masses are added for the galaxies
without measured $R_e$ and $\sigma_e$.  As is evident in Fig.~\ref{sm}, the extra points are particularly
valuable for the lower-luminosity galaxies.

\section{Discussion}

\subsection{Population Scaling Relations}

We can gain some more understanding of why the relation shown in 
Fig.~\ref{genfit} between cluster population and $(R_e \sigma_e)$ works by
looking further at the scaling relations among galaxy mass, luminosity,
size, and velocity dispersion.  For the giant ellipticals ($L> 10^{10} L_{\odot}$),
direct fits of each of $R_e$ and $\sigma_e$ versus luminosity 
give $R_e \sim L^{0.66\pm0.033}$ and $\sigma_e \sim L^{0.285 \pm 0.017}$.  Combining these then predicts
$(R_e \sigma_e) \sim L^{0.95 \pm 0.04} \sim M_{dyn}^{0.79 \pm 0.04} \sim N_{GC}^{0.76 \pm 0.04}$, 
using the other correlations in Table 2 to translate from $L$ to $M_{dyn}$ and then to $N_{GC}$.
Inverting the result then gives $N_{GC} \sim (R_e \sigma_e)^{1.31 \pm 0.07}$,
which closely matches what we obtain from the direct solution in Fig.~\ref{genfit} and Table 2.

The dwarf galaxies obey somewhat different scalings among size, dispersion, and luminosity, namely
$R_e \sim L^{0.26\pm0.06}$ and $\sigma_e \sim L^{0.32 \pm 0.06}$.  
Combining these leads to $(R_e \sigma_e) \sim M_{dyn}^{0.48 \pm 0.07}$.
However, this shallower trend is partly compensated by the shallower dependence of cluster
population on mass for the dwarfs (Table 2), $N_{GC} \sim M^{0.37 \pm 0.09} \sim (R_e \sigma_e)^{0.78 \pm 0.17}$.  The luminosity range of the dwarf E's is small
enough that they can accommodate a relatively wide range of slopes, permitting a single linear relation across the
entire range from dwarf spheroidals to supergiants to be a workable representation.
Said differently, the GC population of a galaxy can be
seen as another outcome of the Fundamental Plane for early-type galaxies.\footnote{If we assume
more generally that $N_{GC} \sim R_e^a \sigma_e^b$, it can be
shown from the scalings between $R_e, \sigma_e$, and $L$ listed above that any pair
of exponents where $a \simeq 6.9 b$ will work for both the giants and dwarfs.  However, the
combination $N \sim (R \sigma)^{1.3}$ has the strong advantage of simplicity and reproduces
the actual data well.}

\subsection{Mass Fractions and Formation Efficiency Parameters}

Discussion of the specific frequency and specific mass $S_N$ and $S_M$ quickly
leads to the question of GC formation efficiency -- the original reason
for defining these ratios.  A summary of the thinking during the early (pre-$\Lambda$CDM) literature
can be found in \citet{har01}.  A more recent view with growing evidence
is that the GCs, which are compact stellar subsystems emerging from the densest and most massive sites of star formation,
could be the objects that form earliest in any starburst, followed by the bulk of the field stars and the lower-mass
star clusters that soon dissolve into the field.  
First ideas along these lines were explored by \citet{bla97, mcl99, kav99}, and \citet{har02},
pointing to the possibility of
a universal GC formation efficiency \emph{per unit baryonic mass} including both stars and gas.
If the later rounds of star formation {\sl after} GC formation
are truncated or severely reduced by any combination of external or internal quenching, then the resulting
$S_N$ or $S_M$ observed long after the fact is a marker of how well the quenching worked.  High$-S_N$ systems thus
would be \emph{field-star deficient} (not ``cluster-rich'') because the star formation
was prevented from running to completion.  This interpretation has been developed further
by P08, G10, and \citet{spi10}; in particular, P08 estimate 
quantitatively that in dwarf galaxies the peak star formation epoch lagged the peak GC formation epoch by
$350 - 600$ Myr.

Different mechanisms for shutting down star formation will operate at the opposite extremes of the
galaxy mass range.  For the dwarfs with their small potential wells, internally driven feedback including starburst winds,
photoionization, and ultimately SNe may eject a large fraction of the gas, while external quenching
from tidal stripping of gas or external UV fields can also reduce the star formation efficiency (SFE).
For the most massive galaxies, AGN feedback and virial shock heating of infalling gas will lower
the SFE \citep{dek86, dek03,dek06,bow06}. 

The maximum SFE should then happen for intermediate-mass galaxies
which are massive enough to hold their star-forming gas against SN and starburst winds, but 
in which AGN feedback or shock heating are not intense enough to have much effect.  Much recent literature
has addressed this issue by tracing the change in the ratio $M(halo)/M(baryon)$ versus galaxy size,
which displays the same U-shaped distribution that we see in the $S_N$ and $S_M$ curves.
To mention only two recent studies, \citet{sha06} identify SFE(max) at a stellar
mass $M_{\star} \simeq 6.3 \times 10^{10} M_{\odot}$ or $M(halo) \simeq 10^{12} M_{\odot}$, while from 
a combination of direct stellar mass and halo mass observational determinations
\citet{lea12} find the SFE maximum to be at $M_{\star} \simeq 4 \times 10^{10} M_{\odot}$.

Viewed in this light, the {\sl minimum} of the $S_M$ or $S_N$ distributions becomes perhaps
the most interesting region of those diagrams.  The baseline ratios $S_N \simeq 1$ and $S_M \simeq 0.1$
essentially tell us, \emph{in the galaxies where star formation was globally the most efficient}, what
mass fraction went into the dense compact systems that we {\sl now} see surviving as globular clusters.
That is, \emph{these baseline values reveal what the ``natural'' GC formation 
efficiency is when any disruption or quenching of star formation is at its least important.}

This present-day mass fraction is, however, only a lower limit to the value at the time of formation,
because (a) many low-mass or low-density star clusters are disrupted or dissolved over the subsequent
Hubble time, and (b) even the surviving ones that started
as the densest, most massive clumps have lost a large fraction of their initial mass through a combination
of early rapid mass loss (expulsion of residual gas, and SNe and stellar winds from massive stars) and later
dynamical erosion.  The current literature \citep{tre07,mcl08,ves10} 
indicates that the individual globular clusters that have survived to the present should have been
$\sim$ 10 times more massive when they were protoclusters than they are now.  The conclusion we draw from these arguments is that 
the surviving GCs 
represented at least 1 percent of the star formation mass fraction in the maximally efficient intermediate-mass galaxies.
In either dwarf or supergiant galaxies, however, where $S_M$ may be an order of magnitude higher,
the GCs we see today 
could -- as protoclusters -- have taken as much as 10\% of the gas that successfully formed stars.
And these mass fractions must only be lower limits to the amounts of star-forming gas
that went into young star clusters, after accounting for the clusters that did not survive to the present.

The interpretation that $M_{GCS}$ is driven by the amount of gas mass initially present in the
galaxy's potential well (and not the gas mass that was actually consumed in star formation) then
raises the possibility that $M_{GCS}$ should be more nearly proportional to $M_{halo}$, or the
total depth of the galaxy potential well.  This direction has been
explored by \citet{bla97,mcl99,spi08}, P08, S09, and G10.
Following the notation of G10, we denote $\eta \equiv M_{GCS}/M_{halo}$.
For the dwarf galaxies in particular (discussed at greater length by P08 and G10),
our derivation that $M_{GCS} \sim M_{dyn}^{0.46 \pm 0.10}$ is in excellent agreement with
the scaling $M_{halo} \sim M_{\star}^{0.46}$ obtained by \citet{lea12} if we assume that
$M_{GCS} \propto M_{halo}$ and also $M_{\star} \simeq M_{dyn}$ as above.

To derive the ``absolute'' GC formation efficiency parameter
$\eta$ from our new sample of galaxies, we need to adopt a stellar-to-halo mass conversion relation 
$M_{halo}  = f(M_{\star})$, or else its inverse.  
By hypothesis we also simply use $M_{\star} \simeq M_{dyn}$ and $\eta = const$  as mentioned above.
These steps then directly link $M_{halo}$ to $M_{GCS}$ and $M_{\star}$, and the assumed value of $\eta$ can be
varied until a match is achieved with our data.
Our specific approach is to fix $\eta$ by requiring
the resulting $S_M$ vs.~$M_{\star}$ curve to pass through the baseline $S_M \simeq 0.1$ for the 
intermediate-mass galaxies.  

\begin{figure}
\plotone{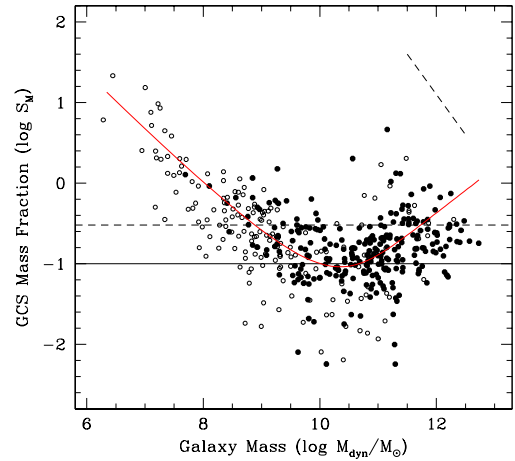}
\caption{Specific mass $S_M = 100 (M_{GCS}/M_{dyn})$ versus host galaxy mass.
\emph{Solid dots} are dynamical masses $M_{dyn}$ calculated from $(R_e, \sigma_e)$,
while \emph{open circles} are masses estimated from $L_V$ and Eq.~3, for galaxies
without measurements of $R_e$ and $\sigma_e$.
The solid diagonal line at upper right shows the effect of changing $M_{dyn}$ by 1.0 dex
(an increase in $M_{dyn}$ yields a proportionate decrease in $S_M \sim M_{GCS}/M_{dyn}$).
The horizontal line at $S_M = 0.1$ is the approximate average level reached for the 
intermediate-luminosity galaxies where star formation is maximally efficient.
80\% of all galaxies fall below the dashed horizontal line at $S_M = 0.3$ (see text).
The superimposed curve is the interpolation model fit discussed in Section 4.2.
}
\label{sm}
\end{figure}

One such conversion between $M_{\star}$ and $M_{halo}$  is given by \citet{beh10} and \citet{lea12}, derived from
a combination of methods for measuring $M_{halo}$ and $M_{\star}$ over a wide range of luminosity
regimes (see Eq.~ 24 from Behroozi or Eq.~13 from Leauthaud).  Their empirical model shows a clear rise
in the ratio $M_{halo}/M_{\star}$ at the low-mass and high-mass ends with a minimum at
intermediate galaxies.   
However, their model function and parameters give a curve for $S_M$
that is too steep at each end to be satisfactory in detail for our purposes.  A flexible and simpler 
conversion relation from \citet{yan08} and also used by S09 is
\begin{equation}
M_{\star} \, = \, M_0 { {(M_{halo}/M_1)^{\alpha + \beta}} \over {(1 + M_{halo}/M_1)^{\beta} }} \, .
\end{equation}
To use this, we assume a value for $M_{halo}$, which determines $M_{\star}$ (which by hypothesis equals our $M_{dyn}$).
Finally $S_M = 100 \cdot M_{GCS}/M_{\star}$, which then defines a point on Fig.~\ref{sm}.
Repeating for a wide range of $M_{halo}$ then defines a complete $S_M$ vs. $M_{dyn}$ curve.

An illustrative fit of this model to the GCS data is shown as the solid curve in Fig.~\ref{sm}.  This curve uses
the parameters 
$\eta = 6.0 \times 10^{-5}$, log $M_0$ = 9.98, and log $M_1$ = 10.7 along
with exponents $\alpha = 0.64, \beta = 2.88$.\footnote{The 
powerlaw-like slopes of the model curve at the high and low mass ends are quite sensitive to the choices
of ($\alpha, \beta$), and the values 
we find to give a good fit are slightly different from the ones used by S09.  Again, they used
a different, simpler prescription for $\langle M_{GC} \rangle$ and thus $M_{GCS}$; a smaller GCS dataset; and a different
prescription for computing galaxy masses.}
At either the low-mass or high-mass end of the scale, the $S_M$ curve asymptotically approaches
a simple power law.  At the high-mass end, where $(M_{halo}/M_1)$ becomes large, it can quickly
be shown that $S_M \rightarrow const \cdot M_{dyn}^{(1 - \alpha)/\alpha}$, which for our fitted
value $\alpha \simeq 0.64$ gives $S_M\sim M_{dyn}^{0.563}$.
At the low-mass end where $(M_{halo}/M_1)$ is small, then 
$S_M \rightarrow const \cdot M_{dyn}^{-1 + 1/(\alpha+\beta)}$, which for $\beta = 2.88$ gives
$S_M \sim M_{dyn}^{-0.716}$.

Our estimated $\eta \simeq 6 \times 10^{-5}$ represents the absolute efficiency of GC formation.
By comparison, through different combinations of methods for deriving the various masses and luminosities,
G10 find a mean $\langle \eta \rangle \simeq  6 \times 10^{-5}$, while S09
find $\eta \simeq 7 \times 10^{-5}$.  The agreement among these discussions 
is well within the scatter we can expect given the
different assumptions for the definition of $S_M$ and the 
methods for finding galaxy luminosities, stellar masses, and dark-halo masses.

We have suggested in the discussion above that the galaxy mass where $S_M$ reaches a minimum
represents the level where star formation efficiency was the highest.
The $S_M$ interpolation curve in Fig.~\ref{sm} reaches a minimum at $M_{gal} \simeq 2.5 \times 10^{10} M_{\odot}$.
Within the accuracy permitted by the scatter around the curve, 
this minimum point is strikingly similar to the galaxy-based estimates of 
$M_{gal} \sim 5 \times 10^{10} M_{\odot}$ mentioned above for the point of maximum SFE, 
and emerges from quite a different line of argument.
Yet another piece of information in line with these results is recent evidence 
\citep{spo09,spo10,tor10} that 
the maximum \emph{metallicity gradient} within galaxies occurs near $M_{gal} \sim 3 \times 10^{10} M_{\odot}$,
as would be the case for galaxies that have had minimal influences from mergers and feedback during their
primary star-forming stages.

Along with the GCS catalog itself, we view Fig.~\ref{sm} and the discussion above
as the most important result of this paper.  The relative number of globular clusters in a galaxy as
measured by either $S_N$ or $S_M$ differs considerably from one system to another, but still follows a systematic trend
that can be matched by a single, constant ratio $\eta = M_{GCS}/M_{halo} \simeq 6 \times 10^{-5}$.
Our results strongly support recent work (S09, G10) that the globular cluster system 
is a remarkably simple proxy for the most fundamental characteristic of
a galaxy, namely its dark-matter potential well.

The hypothesis $M_{GCS} \propto M_{halo}$ can explain the basic shape of the $S_M$ distribution
with its characteristic rise at extreme low and high luminosities, but it does not
address the {\sl scatter} that we see in any given mass range.    
Quantitatively this scatter is $\pm 0.25$ dex at any point on the mean curve,
or slightly less than a factor of two in $S_M$. 
We should expect four generic sources of scatter:
\begin{itemize}
\item{} Random measurement uncertainties in $N_{GC}$ and thus $M_{GCS}$.
The raw uncertainties or even the statistical
variance in $N_{GC}$, as discussed above,
may be up to factors of 2 depending on the galaxy and are therefore 
probably the dominant source of the observed scatter.  
\item{} Random uncertainties in the quantities that determine $M_{dyn}$, namely
$R_e$ and $\sigma_e$.  These behave differently from random scatter in $N_{GC}$,
since $S_M \sim M_{GCS} / M_{dyn}$, and thus $S_M$ and $M_{dyn}$ are not
independent quantities.  
Any error in $M_{dyn}$ would generate
an equal, inverse change in $S_M$ and would shift points along
a diagonal line in Fig.~\ref{sm}.  If this was, however, an important source of the observed
scatter then it should be most visible at the high-mass end where the 
$S_M$ curve is nearly perpendicular to that error line.  In practice we see
rather similar amounts of scatter over a wide range of masses,
consistent with the expectation from the raw observations that
$R_e$ and $\sigma_e$ are uncertain to $\lesssim 10$\% \citep[see][]{har13}.  
\item{} Intrinsic differences in GC formation efficiency between galaxies of
similar type and mass.  These differences certainly exist (compare the classic
well-studied 
cases of the Virgo giants M87 and M49, which have similar luminosities but GC
populations different by almost a factor of 3), but it is harder to make any general
statements about the amount of such ``cosmic scatter'' at this stage.
A large part of the scatter may be the result of environment:  for example, P08 address this question
in detail for dwarfs, and find evidence that dE galaxies near dominant giants
are more likely to have higher$-S_N$ GCSs. 
By hypothesis these galaxies
may have benefitted from being in deeper halo potential wells which increased
massive star cluster formation.  
At the opposite end of the environment scale, \citet{cho12} find that E galaxies in very
isolated environments have quite low specific frequencies in the range normally associated
with spirals or S0's.
\item{} The stochastic effects of different individual merging and star-forming histories, which are hard to recover in
full detail long after the fact.  These must also play a role in generating galaxy-to-galaxy differences.
Conversely, differences in dynamical GC destruction 
interior to a galaxy should not be a major factor, since internal GC erosion rates
should be similar \emph{for galaxies of the same type and luminosity}.  
Going further into these intriguing questions
is beyond the scope of our paper.
\end{itemize}

Lastly, although some dwarfs and supergiants stand out as having exceptionally high $S_M$,
it is perhaps worth noting as well that many dwarfs and giants 
have $S_M$ or $S_N$ values that are quite similar
to those of the intermediate-luminosity galaxies in the ``baseline'' middle range.
The large size of our new catalog allows this feature of the distribution to stand
out more clearly than before.  
Quantitatively, fully 80\% of the \emph{entire} sample in Fig.~\ref{sm} falls within
$S_M < 0.3$ (below the dashed line in Fig.~\ref{sm}), and 64\% fall within $S_M < 0.2$.  The median of the whole sample
(less sensitive to outliers than the mean) is at $\tilde S_M = 0.153$.
In terms of specific frequency, 64\% of the entire catalog falls below $S_N =3$ and
the median is at $\tilde S_N = 2.07$.

It seems that no single starting
point for GC formation may work equally well for every galaxy.  If GCs form extremely early
and the later field-star formation is quenched or disrupted, then the result will be
a cluster-rich galaxy.  Contrarily, if there is little difference between GC formation times and
field-star formation times, then a lower$-S_N$ GC system will result.
Therefore we suggest that the many dwarfs and giants with lower 
specific frequencies could be ones in which 
the GC and field-star formation rates versus time were the same as in the
intermediate galaxies.

A next step in understandng the link between GC formation and
their host galaxies will be to explore more thoroughly
the relation with cluster metallicity
(color):  what does the correlation of $S_M$ versus galaxy mass
look like for the red and blue GC subgroups?  
Another major question is the role of galaxy environment, and how
much of the scatter around the $S_M$ and $S_N$ relations is driven
by a galaxy's location.  Initial work on these questions has started,
but extensions to much bigger samples will be valuable.

\section{Summary}

We summarize the results of our discussion as follows:

\begin{itemize}
\item{}A new catalog of 422 galaxies with published measurements of their
globular cluster systems is presented along with a source bibliography.  
This list, based on a literature survey to the end of 2012, contains 
248 ellipticals (dwarfs and giants), 93 S0's, and 81 spirals and
irregulars.
\item{} Total GC population $N_{GC}$ increases monotonically with either
host galaxy luminosity or baryonic mass, but not in a simple linear way.  In agreement
with other recent studies but now based on a larger sample, we find that the GC specific
frequency and specific mass follow a U-shaped trend, with very high $S_N$ at either very low
or very high luminosity, but reaching a well defined mean value 
$S_N = 1$ and $S_M = 0.1$ in the range 
$M_{dyn} \sim 2-3 \times 10^{10} M_{\odot}$.  This trend can be understood as the result
of the different kinds of feedback operating during galaxy formation:  
for the low-mass dwarfs, field-star formation is inhibited by radiative feedback,
gas ejection, and externally driven damping before it can run to completion; while for giants, early AGN
activity and virial heating inhibit field-star formation after the GCs have
formed.  At intermediate galaxy mass, neither kind of feedback is as
important and so these galaxies can form stars the most efficiently.
Thus along with P08, G10 we identify these minimum $S_N, S_M$ values as the 
{\sl baseline normal} for star formation minimally damped by feedback or external quenching.
\item{} High-$S_N$ galaxies such as the extreme dwarfs or supergiants may be ones
in which the GC formation epoch preceded the bulk of field-star formation and was
therefore less affected by feedback and quenching processes.  
\item{} Previous recent studies including \citet{pen08,spi08,spi09,geo10} 
have explored the proposal that GC population size (or more importantly,
the mass $M_{GCS}$) is directly proportional to the host galaxy halo mass
$M_{halo}$.  Our work adds support to this interpretation.  We find that 
a single constant ratio $\eta \equiv M_{GCS}/M_{halo} = 6 \times 10^{-5}$
is capable of reproducing the systematic trend of specific mass $S_M$ versus
galaxy mass, over the entire range of galaxy sizes and masses.
The galaxy-to-galaxy scatter anywhere around this relation is typically a factor of two.
\item{} We find that GC population size can also be accurately predicted
by a simple product of galaxy effective radius and velocity dispersion, as
$N_{GC} \sim (R_e \sigma_e)^{1.3}$.  
The residual scatter is $\pm0.32$ dex, making it competitive with any other
proposed correlation.  
We show that this relation can be roughly understood from previously known 
Fundamental Plane scaling
relations among galaxy luminosity, mass, and scale size.
\end{itemize}

\acknowledgments
This work makes use of data products from the Two Micron All-Sky Survey
which is a joint project of the University of Massachusetts and the Infrared
Processing and Analysis Center/California Institute of Technology, funded
by the National Aeronautics and Space Administration and the National Science
Foundation.
This work was supported in part by the Natural Sciences and Engineering
Research Council of Canada through research grants to WEH, and by McMaster
University through partial summer student salary to MA.  GLHH wishes to
thank ESO/Garching for a visiting scientist fellowship, where the first steps toward building this catalog
were carried out.  We thank the anonymous referee for suggestions and comments
that improved the presentation of this paper.


\clearpage






\begin{deluxetable}{llccllcc}
\tabletypesize{\footnotesize}
\tablecaption{Correlation Solutions \label{solutions}}
\tablewidth{0pt}
\tablehead{
\colhead{(x, y)} & \colhead{Galaxy Type} & \colhead{N(sample)} & \colhead{Mean} & \colhead{Zeropoint} & \colhead{Slope} & 
\colhead{Residual} & \colhead{RMS} \\
 & & & $x_0$ & $\alpha$ & $\beta$ & Dispersion & Scatter \\
 & & & & & & $\epsilon_y$ & $\sigma_y$ \\
}

\startdata
(log $L_V$, log $M_{dyn}$) & All Ellipticals & 161 & 10.2 & 10.844 $\pm$ 0.021 & 1.200 $\pm$ 0.021 & 0.24 & 0.27 \\
(log $L_K$, log $M_{dyn}$) & All & 238 & 10.7 & 10.786 $\pm$ 0.017 & 1.085 $\pm$ 0.020 & 0.255& 0.265\\
& $M_K < -22.4$ & 174 & 11.1 & 11.212 $\pm$ 0.019 & 1.260 $\pm$ 0.048 & 0.240& 0.250\\
& $M_K > -22.4$ &  67 & 9.5 & 9.531 $\pm$ 0.033 & 0.765 $\pm$ 0.060 & 0.26 & 0.268\\
(log $M_{dyn}$, log $N_{GC}$) & Luminous E's & 139 & 11.2 & 2.924 $\pm$ 0.028 & 1.035 $\pm$ 0.033 & 0.28 &0.32 \\
& Dwarf E's & 35 & 9.2 & 1.274 $\pm$ 0.045 & 0.365 $\pm$ 0.091 & 0.24 & 0.27 \\
(log $E_b$, log $N_{GC}$) & Luminous E's & 129 & 16.0 & 3.075 $\pm$ 0.027 & 0.735 $\pm$ 0.028 & 0.29 & 0.31 \\
(log $R_e \sigma_e$, log $N_{GC}$) & All Ellipticals & 158& 0.20 & 2.776 $\pm$ 0.025 & 1.290 $\pm$ 0.033 & 0.29 & 0.32 \\
(log $R_e$, log $N_{GC}$) & S/Irr & 60 & 0.4 & 1.582 $\pm$ 0.069 & 0.995 $\pm$ 0.107 & 0.50 & 0.53 \\
(log $M_{dyn}$, log $M_{GCS}$) & Luminous E's & 125 & 11.4 & 8.625 $\pm$ 0.028 & 1.160 $\pm$ 0.044 & 0.29 & 0.31 \\
 & Dwarf E's & 36 & 9.2 & 6.524 $\pm$ 0.046 & 0.460 $\pm$ 0.097 & 0.25 & 0.27 \\
(log $L_V$, log $R_e$) & Luminous E's & 136 & 10.5 & 0.648 $\pm$ 0.018 & 0.660 $\pm$ 0.029 & 0.20 & 0.21 \\
& Dwarf E's & 61 & 8.5 & 0.020 $\pm$ 0.026 & 0.255 $\pm$ 0.054 & 0.20 & 0.21 \\
(log $L_V$, log $\sigma_e$) & Luminous E's & 142 & 10.5 & 2.310 $\pm$ 0.009 & 0.285 $\pm$ 0.017 & 0.09 & 0.11 \\
& Dwarf E's & 36 & 8.7 & 1.566 $\pm$ 0.025 & 0.315 $\pm$ 0.058 & 0.14 & 0.15 \\
(log $M_{dyn}$, log $\langle M_{GC} \rangle$) & All & 242 & 10.7 & $5.402 \pm 0.006$ & $0.098 \pm 0.009$ & 0.028 & 0.086 \\
\enddata

\end{deluxetable}

\end{document}